\documentclass[aps,prl,twocolumn,showpacs,groupedaddress]{revtex4}
\usepackage{graphics}
\usepackage{graphicx}

\begin{document}


\title{Enhanced Transmission of Light and Matter through Nanoapertures
without Assistance of Surface Waves}


\author{S.V. Kukhlevsky}
\affiliation{Department of Physics, University of P\'ecs,
Ifj\'us\'ag u.\ 6, H-7624 P\'ecs, Hungary}


\begin{abstract}
Subwavelength aperture arrays in thin metal films enable enhanced
transmission of light and matter waves [for example, see T.W.
Ebbesen et al., Nature (London) 391, 667 (1998) and E. Moreno et
al., Phys. Rev. Lett. 95, 170406 (2005)]. The phenomenon relies on
resonant excitation of the surface electron or matter waves. We
show another mechanism that provides a great transmission
enhancement not by coupling to the surface waves but by the
interference of diffracted evanescent waves in the far-field zone.
Verification of the mechanism is presented by comparison with
recently published data.
\end{abstract}


\maketitle

%
\maketitle
%
%
The scattering of waves by apertures is one of the basic phenomena
in the wave physics. The most remarkable feature of the light
scattering by subwavelength apertures in a metal screen is
enhancement of the light by excitation of electron waves in the
metal. Since the observation of enhanced transmission of light
through a 2D array of subwavelength metal nanoholes \cite{Ebb},
the phenomenon attracts increasing interest of researchers because
of its potential for applications in nanooptics and nanophotonics
\cite{Barn}. Recently, the enhanced transmission through
subwavelength apertures was predicted also for matter waves
\cite{More}. The enhancement of light is a process that can
include resonant excitation and interference of surface plasmons
\cite{More,Schr,Sobn,Port}, Fabry-Perot-like intraslit modes
\cite{Asti,Taka,Lala,Barb}, and evanescent electromagnetic waves
at the metal surface \cite{Leze}. In the case of thin screens
whose thickness are too small to support the intraslit resonance,
the extraordinary transmission is caused by the excitation of
surface plasmons or their matter-wave analog, surface matter waves
\cite{More,Schr,Sobn,Port}. In this Letter, we show another
mechanism that provides a great transmission enhancement not by
coupling to the surface electron or matter waves but by the
interference of diffracted evanescent waves in the far-field zone.

The transmission enhancement without assistant of surface waves
can be explained in terms of the following theoretical
formulation. We first consider the transmission of light through a
structure that is similar but simpler than an array of holes,
namely an array of parallel subwavelength slits. The structure
consists of a thin metal screen with the slits separated by many
wavelength. To exclude the plasmons from our model, the metal is
considered to be a perfect conductor. Such a metal is described by
the classic Drude model for which the plasmon frequency tends
towards infinity. Owing to the great  distance between the slits,
the electromagnetic field at one slit is assumed to be independent
from other slits. The transmission of the slit array is determined
by calculating the light power in the far-field diffraction zone.
The waves diffracted by each of the independent slits are found by
using the Neerhoff and Mur approach, which uses a Green's function
formalism for a rigorous numerical solution of Maxwell's equations
for a single, isolated slit \cite{Neer,Harr,Betz}. The
calculations show a big, up to 5 times, resonant transmission
enhancement near to the Fabry-Perot wavelengths determined by an
array period. To clarify the numerical result, we then present an
intuitively transparent analytical model, which quantitatively
explains the resonant enhancement in terms of the far-field-zone
interference of the evanescent waves produced by the independent
slits. The model predicts the $\sim$5-times (resonant) and
$\sim$1000-times (nonresonant) enhancements for both the light and
matter waves passing through a perforated metallic or dielectric
screen, independently on the apertures shape. Verification of the
analytical formulae by comparison with data published in the
literature supports these predictions. The Wood anomalies in
transmission spectra of optical gratings, a long standing problem
in optics \cite{Hess}, follow naturally from interference
properties of the model.

Let us first investigate the light transmission by using the
rigorous model. The model considers an array of $M$ independent
slits of width $2a$ and period $\Lambda$ in a screen of thickness
$b\ll\lambda$. The screen placed in vacuum is illuminated by a
normally incident TM-polarized wave with wavelength
$\lambda=2{\pi}c/\omega=2\pi/k$. The magnetic field of the wave
${\vec{H}}(x,y,z,t)=U(x){\exp}(-i(kz+\omega{t})){\vec{e}}_y$ is
assumed to be time harmonic and constant in the $y$ direction. The
transmission of the slit array is determined by calculating all
the light power $P(\lambda)$ radiated into the far-field
diffraction zone, $x{\in}[-\infty,\infty]$ at the distance
$z\gg{\lambda}$ from the screen. The total per-slit transmission
coefficient, which represents the per-slit enhancement in
transmission achieved by taking a single, isolated slit and
placing it in an $M$-slit array, is then found by using an
equation $T_M(\lambda)=P(\lambda)/MP_1$, where $P_1$ is the power
radiated by a single slit. Figure 1 shows the transmission
coefficient $T_M(\lambda)$, in the spectral region 500-2000 nm,
calculated for the array parameters: $a=100$ nm, $\Lambda=1800$
nm, and $b=5\times 10^{-3}\lambda_{max}$.
\begin{figure}
\begin{center}
\includegraphics[keepaspectratio, width=1\columnwidth]{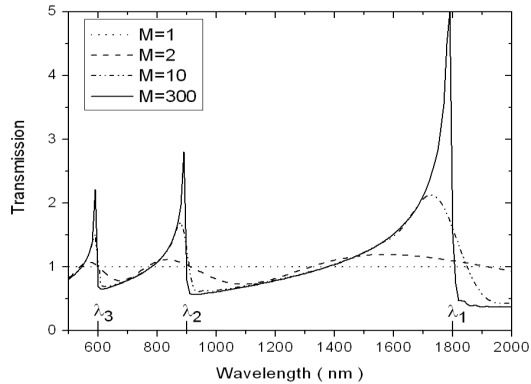}
\end{center}
\caption{The per-slit transmission $T_M(\lambda)$ of an array of
independent slits versus the wavelength for different number $M$
of slits. There are three Fabry-Perot resonances at the
wavelenghts $\lambda_n{\approx}\Lambda/n$, $n$=1, 2 and 3.}
\label{fig:Fig1}
\end{figure}
The transmitted power was computed by integrating the total energy
flux at the distance $z$ = 1 mm over the detector region of width
$\Delta{x}$ = 20 mm. The transmission spectra $T_M(\lambda)$ is
shown for different values of $M$. We notice that the spectra
$T_M(\lambda)$ is periodically modulated, as a function of
wavelength, below and above a level defined by the transmission
$T_1(\lambda)=1$ of one isolated slit. As $M$ is increased from 2
to 10, the visibility of the modulation fringes increases
approximately from 0.2 to 0.7. The transmission $T_M$ exhibits the
Fabry-Perot like maxima around wavelengths
$\lambda_{n}=\Lambda/n$. The spectral peaks increase with
increasing the number of slits and reach a saturation
($T_M^{max}\approx5$) in amplitude by $M=300$, at
$\lambda\approx{1800}$ nm. The peak widths and the spectral shifts
of the resonances from the Fabry-Perot wavelengths decrease with
increasing the number $M$ of slits. From the data of Fig.~1, one
can understand that enhancement and suppression in the
transmission spectra are the natural properties of an ensemble of
independent subwavelength slits in a thin ($b\ll\lambda$) screen.
The spectral peaks are characterized by asymmetric Fano-like
profiles. Such modulations in the transmission spectra are known
as Wood's anomalies. The minima and maxima correspond to Rayleigh
anomalies and Fano resonances, respectively~\cite{Hess}. The weak
Wood's anomalies are present also in the case of $M=2$, a
classical Young type two-slit system.

The above-presented data is based on calculation of the energy
flux by using the electromagnetic field evaluated numerically. The
transmission enhancement is achieved by taking a single, isolated
slit and placing it in an array. The interference of the waves
diffracted by the independent slits can be considered as a
physical mechanism responsible for the enhancement. To clarify the
numerical results and gain physical insight into the enhancement
mechanism, we have developed an analytical model, which yields
simple formulae for the diffracted fields. For the fields
diffracted by a narrow ($2a\ll\lambda, b\geq0$) slit into the
region $|z|> 2a$, it can be shown that the Neerhoff and Mur model
simplifies to an analytical one. For the magnetic
$\vec{H}=(0,H_y,0)$ and electric $\vec{E}=(E_x,0,E_z)$ fields we
found:
\begin{eqnarray}
{H_y}(x,z)=i{a}DF_0^{1}(k[x^2+z^2]^{1/2}),
\end{eqnarray}
\begin{eqnarray}
E_{x}(x,z)={{-az}{[x^2+z^2]^{-1/2}}}D
F_1^{1}(k[x^2+z^2]^{1/2}),
\end{eqnarray}
and
\begin{eqnarray}
E_{z}(x,z)={{ax}{[x^2+z^2]^{-1/2}}}D
F_1^{1}(k[x^2+z^2]^{1/2}),
\end{eqnarray}
where
\begin{eqnarray}
\label{sz:D:def} D=4k^{-1}[[\exp(ikb)(aA-k)]^{2}-(aA+k)^2]^{-1}
\end{eqnarray}
and
\begin{eqnarray}
\label{sz:A:def}
A=F_0^{1}(ka)+\frac{\pi}{2}[\bar{F}_{0}(ka)F_1^{1}(ka)
+\bar{F}_{1}(ka)F_0^{1}(ka)].
\end{eqnarray}
Here, $F_1^{1}$, $F_0^{1}$, $\bar{F}_{0}$ and $\bar{F}_{1}$ are
the Hankel and Struve functions, respectively. The fields are
spatially nonuniform, in contrast to a common opinion that a
subwavelength aperture diffracts light in all directions uniformly
\cite{Lez}. The fields produced by an array of $M$ independent
slits are given by
$\vec{E}(x,z)=\sum_{m=1}^{M}\vec{E}_{m}(x+m\Lambda,z)$ and
$\vec{H}(x,z)=\sum_{m=1}^{M}\vec{H}_{m}(x+m\Lambda,z)$, where
$\vec{E}_{m}$ and $\vec{H}_{m}$ are the fields of an $m$-th beam
generated by the respective slit. As an example, Fig.~2(a)
compares the far-field distributions calculated by the analytical
formulae (1-5) to that obtained by the rigorous model. We notice
that the distributions are undistinguishable.
\begin{figure}
\begin{center}
\includegraphics[keepaspectratio, width=0.9\columnwidth]{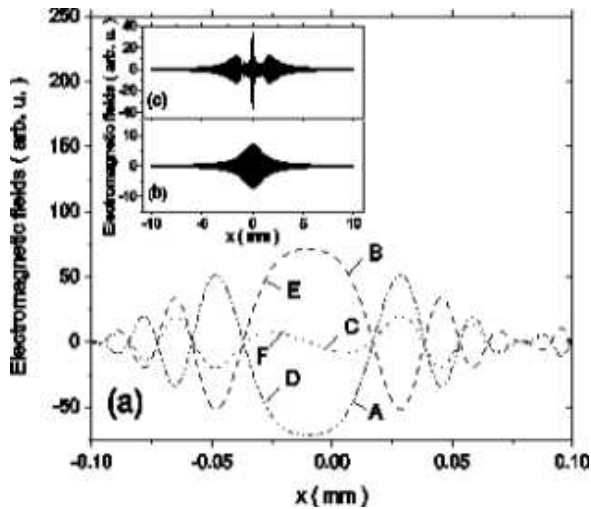}
\end{center}
\caption{Electromagnetic fields in the far-field zone. (a) The
fields Re($E_x(x)$) ($A$ and $D$), Re($H_y(x)$) ($B$ and $E$), and
Re($10E_z(x)$) ($C$ and $F$) calculated for $M$ = 10 and $\lambda$
= 1600~nm. The curves $A$, $B$, and $C$: rigorous model; curves
$D$, $E$, and $F$: analytical model. (b) Re($E_x(x)$) for $M$=1:
analytical model. (c) Re($E_x(x)$) for $M$=5: analytical model.}
\label{fig:Fig2}
\end{figure}
Thus, the analytical model not only supports results of our
rigorous model, but presents an intuitively transparent
explanation of the enhancement in terms of the interference of the
fields produced by the multi-beam source. The array-induced
decrease of the central beam divergence (Figs.~2(b) and 2(c)) is
relevant to the beaming light \cite{Mart}, and the nondiffractive
light and matter waves \cite{Kuk}.

The analytical model accurately describes the fields $\vec{E}$ and
$\vec{H}$, but what is about the transmission coefficient? The
field power $P$, which determines the coefficient $T_M$, is found
by integrating the energy flux
$|\vec{S}|=|\vec{E}\times\vec{H}^*+\vec{E}^*\times\vec{H}|$. Thus,
the model accurately predicts also the light transmission. We now
consider the predictions in light of the key observations
published in the literature for the two fundamental systems of
wave optics, the single-slit and two-slit systems. The major
features of a single-slit system are the intraslit resonances and
the spectral shifts of the resonances from the Fabry-Perot
wavelengths \cite{Taka}. In agreement with the predictions
\cite{Taka}, the formula (4) shows that the transmission $T$ =
$P/P_0$ = $(a/k)[$Re$(D)]^{2}+[$Im$(D)]^2$ exhibits Fabry-Perot
like maxima around wavelengths $\lambda_{n}=2b/n$, where $P_0$ is
the power impinging on the slit opening. The enhancement and
spectral shifts are explained by the wavelength dependent terms in
the denominator of Eq. (4). The enhancement
($T(\lambda_1){\approx}b/{\pi}a$ \cite{Kuk}) is in contrast to the
attenuation predicted by the model \cite{Taka}. The Young type
two-slit configuration is characterized by a sinusoidal modulation
of the transmission spectra $T_2(\lambda)$ \cite{Scho,Lalan}. The
modulation period is inversely proportional to the slit separation
$\Lambda$. The visibility $V$ of the fringes is of order 0.2,
independently of the slit separation. In our model, the
transmission is given by $T_{2}{\sim}{\int}
[F_1^1(x_1)[iF_0^1(x_1)]^{*}+F_1^1 (x_2)^{*}iF_0^1(x_2)]dx$, where
$x_1=x$ and $x_2=x+\Lambda$. The high-frequency modulations with
the sideband-frequency $f_s(\Lambda)$
${\approx}f_1(\lambda)+{f_2(\Lambda,\lambda)}{\sim}1/{\Lambda}$
(Figs. 1 and 3) are produced as in a classic heterodyne system by
mixing two waves having different spatial frequencies, $f_1$ and
$f_2$.
\begin{figure}
\begin{center}
\includegraphics[keepaspectratio, width=0.8\columnwidth]{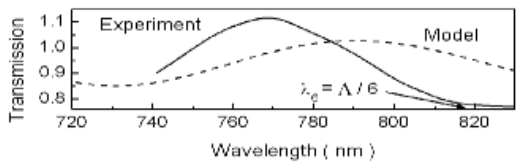}
\end{center}
\caption{The per-slit transmission coefficient $T(\lambda)$ versus
wavelength for the Young type two-slit experiment \cite{Scho}.
Solid curve: experiment; dashed curve: analytical model.
Parameters: $a$ = 100 nm, $\Lambda$ = 4900 nm, and $b$ = 210 nm.}
\label{fig:Fig3}
\end{figure}
Although our model ignores the plasmons, its prediction for the
visibility ($V$ $\approx$ 0.1) of the fringes and the resonant
wavelengths $\lambda_{n}=\Lambda/n$ compare well with the
plasmon-assisted Young's type experiment \cite{Scho} (Fig.~3).
Some difference between the calculated and measured values shows
that the plasmonless and plasmon-assisted resonances can compete
in certain situations. In the case of $b\geq{\lambda/2}$, the
resonances at $\lambda_{n}=\Lambda/n$ can be accompanied by the
intraslit resonances at $\lambda_{n}=2b/n$.

In order to gain physical insight into the mechanism of
plasmonless enhancement in a multi-slit ($M{\geq}2$) system, we
now consider the dependence of the transmission $T_M(\lambda)$ on
the slit separation $\Lambda$. We assume that the slits are
independent also at $\Lambda\rightarrow{0}$.
\begin{figure}
\begin{center}
\includegraphics[keepaspectratio, width=0.8\columnwidth]{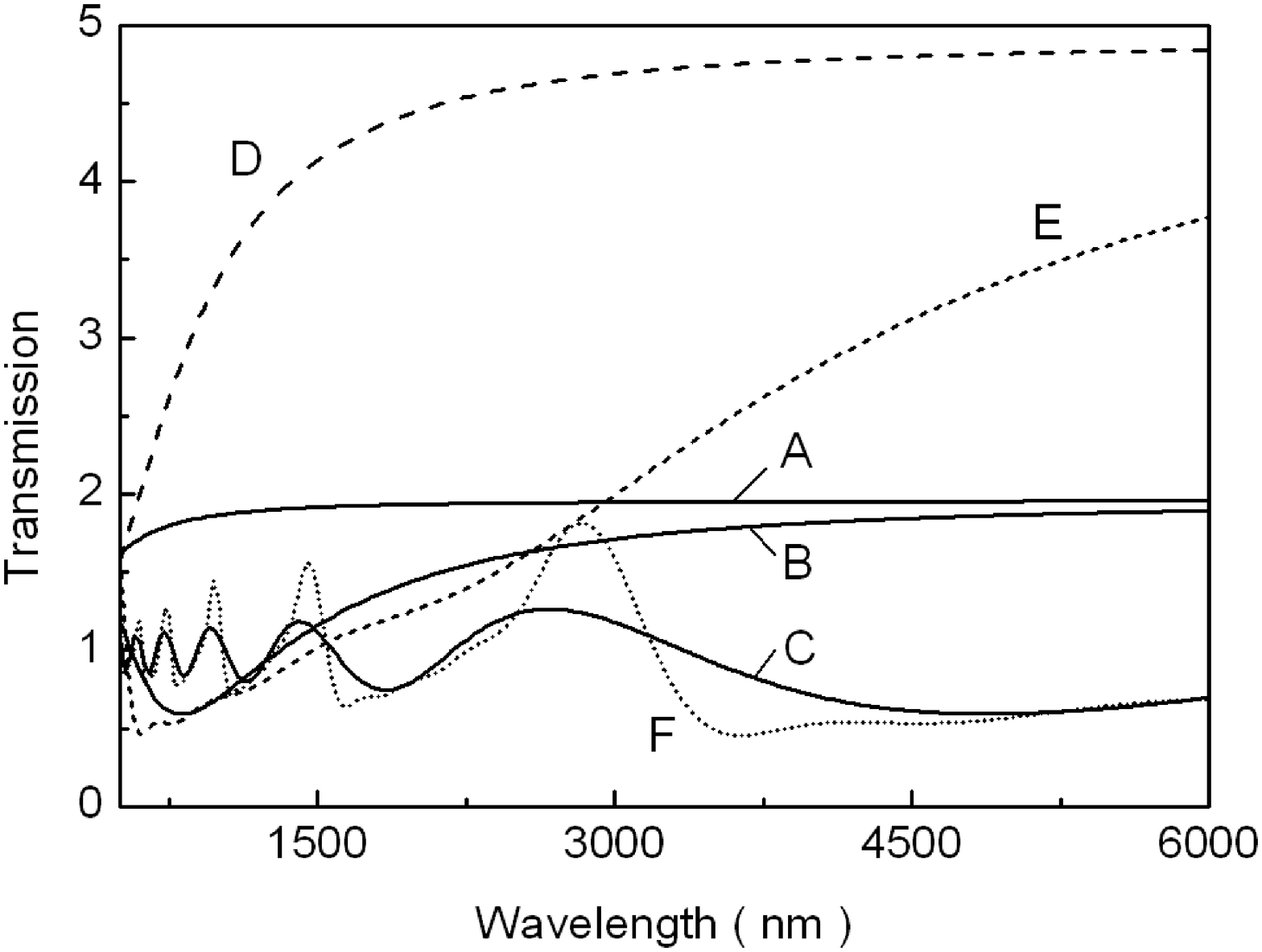}
\end{center}
\caption{The per-slit transmission $T_M(\lambda)$ versus
wavelength for the different values of $\Lambda$ and $M$: (A)
$\Lambda=$ 100 nm, $M=2$; (B) $\Lambda=$ 500 nm, $M=2$; (C)
$\Lambda=$ 3000 nm, $M=2$; (D) $\Lambda=$ 100 nm, $M=5$; (E)
$\Lambda=$ 500 nm, $M=5$; (F) $\Lambda=$ 3000 nm, $M=5$.
Parameters: $a$ = 100 nm and $b$ = 10 nm.  There are two
enhancement regimes at $\Lambda\ll\lambda$ and
$\Lambda{\geq}\lambda$.} \label{fig:Fig3}
\end{figure}
According to the Van Citter-Zernike coherence theorem, a light
source (even incoherent) of radius $r=M(a+{\Lambda})$ produces a
transversally coherent wave at the distance
$z{\leq}{\pi}Rr/\lambda$ in the region of radius $R$. Thus, in the
case of $\Lambda\ll\lambda$, the collective emission of the
ensemble of slits generates the coherent electric and magnetic
fields,
$\vec{E}=\sum_{m=1}^{M}\vec{E}_{m}$exp$(i\varphi_m){\approx}M\vec{E}_{1}$exp$(i\varphi)$
and $\vec{H}{\approx}M\vec{H}_{1}$exp$(i\varphi)$. Consequently,
the maximum power of the emitted light scales with the square of
the number of slits (beams), $P\sim{M^2}$. Therefore, the
transmission ($T_M{\sim}P/M$) grows linearly with the number of
slits, $T_M{\sim}M$. For a given $M$, the function $T_M(\lambda)$
monotonically varies with $\lambda$. Such an enhancement regime
(first regime) is shown in Fig. 4. At appropriate conditions, the
transmission can reach 1000-times enhancement
($M={\lambda}z/\pi{R}(a+\Lambda)$). The transmission enhancement
in the second regime does not require close proximity of the
slits. In the case of $R\geq{\lambda}z/{\pi}r$
($\Lambda{\geq}\lambda$), the beams arrive at the detector with
different phases $\varphi_{m}$. Consequently, the power and
transmission grow slowly with the number of slits (Figs.~1-4).
According to our model, the transmission $T_M$ exhibits the
Fabry-Perot like maxima around wavelengths
$\lambda_{n}=\Lambda/n$. The constructive and destructive
interference of the beams leads respectively to the enhancement
and suppression of the transmission amplitudes as in a classical
heterodyne system.

The analytical model gives not only intuitively transparent
explanation of the plasmonless transmission enhancement, but in
contrast to the previous studies, predicts the enhancement for the
matter waves. Indeed, in our model, the enhancement is based on
coherent excitation of an assemble of slits and constructive
interference of the diffracted waves. The constructive
interference is provided not by coupling between the slits, but by
a geometrically well-defined phase relationship between wave
amplitudes at different lateral locations in the far-field zone.
Thus, the enhancement mechanism depends neither on the nature
(light or matter) of the fields ${\psi}_{m}$ nor on material and
shape of the apertures. For instance, in the first enhancement
regime, the fixed phase correlation leads to the $M$-times
enhancement of the field amplitude,
${\psi}=\sum_{m=1}^M{\psi}_{m}\exp(i\varphi_m){\approx}M{\psi}_{1}\exp(i\varphi)$,
and consequently to the $M$-time transmission enhancement. Due to
Babinet's principle, the model predicts the enhancement also in
the reflection spectra. The destructive interference of the fields
at the detector can lead also to the zero transmission. Indeed,
the interference of the positive (${\psi}$) and negative
($-{\psi}={\psi}\exp(i\varphi+\pi)$) fields produces a field with
the zero amplitude and energy. The value $T$=0 is obtained by
summing up the respective positive and negative energies. Notice
that the amplitudes of the fields ${\psi}_{m}$ can rapidly
decrease with increasing the distances $x$, $y$ and $z$. However,
due to the enhancement and beaming mechanisms (Figs.~1-4), an
array produces a nonevanescent (propagating) wave $\psi$ with low
angle divergence. Such a behavior is in agreement with the
Huygens-Fresnel principle, which considers a propagating wave as a
superposition of secondary spherical waves.

It is worth noting that the presented model is similar in spirit
to the dynamical Bloch-waves diffraction model \cite{Trea}, the
Airy-like model based on the Rayleigh field expansion \cite{Cao},
and especially to a diffracted evanescent wave model~\cite{Leze}.
The difference arises from a fact that the models \cite{Leze,Trea,
Cao} consider the case of $M=\infty$, while predictions of our
model strongly depend on the number $M\neq{\infty}$ of slits. In
addition, our model deals with independent slits, while the models
of Refs. \cite{Leze,Trea,Cao} consider the slits
electromagnetically coupled via the periodic boundary conditions.
There is an evident resemblance also between our model and a Dicke
superradiance model \cite{Dick} of collective emission of an
ensemble of atoms. A quantum reformulation of our model can help
us to understand why a quantum entangled state of photons is
preserved on passage through a hole array \cite{Alte}. The quantum
model will be presented in our next paper. Notice, that the
surface waves can couple the radiation phases of the different
slits, so that they get synchronized, and a collective emission
can release the stored energy as an enhanced radiation. This kind
of enhancement is of different nature compared to our model. The
model  does not require coupling between the slits and does not
contain surface waves.

In conclusion, a rigorous model based on a Green's function
formalism and an analytical model were proposed for description of
the transmission of light through an array of independent slits.
Using the analytical model we showed a physical mechanism that
provides a big transmission enhancement of light or matter waves
not by coupling to the surface waves but by the interference of
diffracted evanescent waves in the far-field zone. The
verification of the analytical formulae by comparison with data
published in the literature supports the predictions. The Wood
anomalies in transmission spectra of optical gratings, a long
standing problem in optics, follows naturally from interference
properties of the model. The analytical formulae can be useful for
experimentalists who develop nanodevices based on transmission and
beaming of light or matter by subwavelengths apertures.
This study was supported by the Hungarian Scientific Research
Foundation (OTKA, Contract No T046811).

%
\end{document}